\documentclass[12pt,times,tight]{aastex62}

\usepackage{amsfonts}
\usepackage{mathrsfs}
\usepackage{bm}
\usepackage{physics}
\usepackage{lipsum}

\def\Gomega{\Gamma_{\omega}}
\def\calL{{\cal L}}
\def\calQ{{\cal Q}}

\def\calA{{\cal A}}

\def\calW{{\cal W}_\Psi}
\def\calWW{{{\cal W}_C}}

\received{***}
\revised{***}
\accepted{***}

\shorttitle{Atlas of 2D Transfer Functions}
\shortauthors{Songsheng et al.}

\begin{document}

\title{\large \bf Kinematic signatures of reverberation mapping of close binaries of supermassive black holes 
in active galactic nuclei. II. Atlas of two-dimensional transfer functions}
\correspondingauthor{Jian-Min Wang}
\email{wangjm@ihep.ac.cn}

\author{Yu-Yang Songsheng}
\affil{Key Laboratory for Particle Astrophysics,
Institute of High Energy Physics,
Chinese Academy of Sciences,
19B Yuquan Road, Beijing 100049, China}
\affil{University of Chinese Academy of Sciences,
19A Yuquan Road, Beijing 100049, China}

\author{Ming Xiao}
\affil{Key Laboratory for Particle Astrophysics,
Institute of High Energy Physics,
Chinese Academy of Sciences,
19B Yuquan Road, Beijing 100049, China}

\author{Jian-Min Wang}
\affil{Key Laboratory for Particle Astrophysics,
Institute of High Energy Physics,
Chinese Academy of Sciences,
19B Yuquan Road, Beijing 100049, China}
\affil{University of Chinese Academy of Sciences,
19A Yuquan Road, Beijing 100049, China}
\affil{National Astronomical Observatories of China,
Chinese Academy of Sciences,
20A Datun Road, Beijing 100020, China}

\author{Luis C. Ho}
\affil{Kavli Institute for Astronomy and Astrophysics, 
Peking University, Beijing 100871, 
China}
\affil{Department of Astronomy, School of Physics, 
Peking University, Beijing 100871, 
China}

\begin{abstract}
Most large galaxies harbor supermassive black holes (SMBHs) in their centers,
and galaxies merge. Consequently, binary SMBHs should be common in galactic 
nuclei.  However, close binaries of SMBH (CB-SMBHs) with sub-parsec separation 
cannot be imaged directly using current facilities. Some indirect signatures, 
such as periodic signals in light curves and double peaks in emission-line 
profile, have been used to find CB-SMBH candidates, but ambiguities still 
exist and no definitive conclusions can be made.
We have recently proposed a new 
method focusing on kinematic signatures that can be derived from 
reverberation mapping of CB-SMBHs, one that offers a promising avenue to  address this 
important problem. In this paper, we calculated models for a wide range of parameters,
but BLRs of two BHs are 
close but still not merged. The purpose of this supplementary paper is to  provide an 
atlas of two-dimensional transfer functions of CB-SMBHs with a wide range of 
orbital and geometrical parameters to aid more efficient identification of CB-SMBH 
candidates in reverberation mapping data.
\end{abstract}



\section{Introduction} \label{sec:intro}
The existence of supermassive black holes (SMBHs) at the center of active galaxies 
was first proposed to explain the ultimate energy source of  quasars 
\citep{Zeldovichi1964, Salpeter1964, Lynden-bell1969}.
Most  galaxies are believed to host quiescent SMBHs,  remnants of their earlier 
active phases \citep{Begelman1980}.  This has since been confirmed from dynamical 
observations of the stars and gas in the Milky Way \citep{Schodel2002} and nearby 
galaxies \citep{Kormendy2013}. Mergers between galaxies are predicted by hierarchical 
structure formation \citep{Lacey1993} in the standard cosmological model, and they 
have been studied statistically in numerous surveys \citep{Patton2002, Lin2004, Conselice2014}.
Binary SMBHs settling in the cores of merged galaxies are thus expected to be quite common, 
with separations and orbits that depend on their  stage of evolution \citep{Begelman1980}.

A number of SMBH binaries with kpc-scale separations have already been 
found \citep{Komossa2003, Bianchi2008, Comerford2009, Comerford2013, Comerford2015, 
Wang2009, Green2010, Koss2011, Liu2011, Liu2018, Fu2015}.
To date, the closest SMBH pair that has been spatially resolved  has a projected separation 
of $7.3$ pc \citep{Rodriguez2006}.  Close binaries of SMBHs (CB-SMBHs) with sub-parsec separation 
are hard to be imaged directly with existing faciltiies.
Alternative indirect methods are needed to confirm their existence observationally.

Periodicity is one of the most distinguishing features of a binary system 
(e.g., \citealt{MacFadyen2008}, \citealt{Roedig2014}, 
\citealt{Farris2015}, \citealt{Shi2015}, \citealt{Bowen2018}).
Periodic signals possibly due to the modulation of orbital motions of SMBH 
binaries have been found in the light curves of a handful of objects, including 
OJ 287 \citep{Sillanpaa1988}, PG~1302$-$102 \citep{Graham2015}, SDSS J0159+0105 \citep{Zheng2016},
NGC 5548 \citep{Li2016} and Ark 120 \citep{Li2017}\footnote{We would point out that there are also 
evidences against the existence of periodic 
signals in some candidates, such as PG~1302$-$102 \citep{Vaughan2016} and OJ 287 \citep{Goyal2018}.}.
For a secure identification of the periodicity  with orbital motion, the monitoring duration should 
be at least 3 times the period \citep{Li2018}. The period of a typical CB-SMBH is
\begin{equation}\label{eq:period}
T = 37.4 M_8^{-1/2} A_{30}^{3/2} \, {\rm yr},
\end{equation}
where $M_8 = M_{\rm tot}/10^8M_{\sun}$ is the total mass of the system, and 
$A_{30} = A/30 {\rm \, light \, days \, (ltd)}$ is the separation between two BHs.
Given the limited time span of the existing monitoring data for most active galactic nuclei (AGNs), 
only binaries with separations around several light days can be detected in this way.
The interaction between the accretion disks \citep{Hayasaki2008} and effects of general relativity 
\citep{Dorazio2015} must be considered, making the theoretical interpretation and 
confirmation nontrivial. 

Another important feature of  CB-SMBHs is the relative velocity between the two BHs, which is about
\begin{equation}
V_{\rm rot}  = 4.13\times 10^3 M_8^{1/2} A_{30}^{-1/2} \, {\rm km \, s^{-1}}.
\end{equation}
When the inclination along the line-of-sight (LOS) and the phase of the rotation are taken into account,
the relative velocity is  sufficiently significant to leave an imprint on the profiles of emission lines, 
such as double-peaked and  asymmetric signatures \citep{Popovic2000, Shen2010, Tsalmantza2011, Popovic2012}.
Shift of line center or shape asymmetry in single-epoch broad line profile have been found in 
objects like SDSS J153636.22+044127.0 \citep{Boroson2009}, NGC 4151 \citep{Bon2012}, and NGC 5548 \citep{Li2016},
making them promising CB-SMBH candidates.
However, the line profile is highly degenerate in its ability to distinguish the intrinsic  dynamics 
of the gas in the broad-line region (BLR).
Complex BLR models can also produce double-peaked or asymmetric broad emission lines \citep{Wang2017}.
These features alone prove the existence of CB-SMBHs, and more information is needed to break the degeneracy.

One possible approach is reverberation mapping (RM), which is used to infer the spatial scale 
of the BLR in time domain \citep{Blandford1982, Peterson1993}.
Although broad-line profiles in the CB-SMBH and disk emitter cases could be similar,
\cite{Shen2010} generally argued that they reverberate differently to the continuum variations. 
We note that velocity-resolved RM data have one more dimension than a line profile,
reflecting the velocity distribution of the ionized gas on different spatial scales,
and so can be used to probe the gravitational potential around the central BH.
Since the gravitational potential around a CB-SMBH is much different than that of a single SMBH,
RM data of CB-SMBHs must contain some distinctive kinematic signatures \citep{Wang2018}.
\citet{Wang2018} provided semi-analytical formulae for two-dimensional (2D) transfer functions (TFs) 
of emission lines to continuum in typical CB-SMBH models,  and several examples were
 illustrated to show the peculiarities of CB-SMBHs in contrast with single SMBHs.
We are conducting a long-term project of Monitoring AGNs with H$\beta$ Asymmetry (MAHA) using the 
Wyoming Infrared Observatory (WIRO) 2.3m
telescope to hunt for CB-SMBHs in the local Universe \citep{Du2018b, Brotherton2019}.
Some indications of CB-SMBHs have been found in the 2D TFs of Mrk 6 and Akn 120, but more data 
are needed for further confirmation.

We also note that the recently deployed GRAVITY instrument on the Very Large Telescope Interferometer 
(VLTI) successfully spatially resolve the BLR in 3C 273  through the $S$-shaped differential phase 
curve (DPC) of the Pa$\alpha$ emission line \citep{Sturm2018}.
DPCs of binary BLRs do possess distinguishing features compared to those from a single BLR due to 
the separation of the two BLRs and the orbital velocity of the system.  These differences can be 
exploited to reliably identify  CB-SMBHs  \citep{Songsheng2019}.
RM campaigns can provide not only CB-SMBH candidates for GRAVITY observations, but they also can 
be analyzed jointly with GRAVITY data to reduce the uncertainties of the orbital parameters and 
provide directly the cosmic distance to the BH  \citep{Wang2019}.

As the MAHA project and other RM campaigns (such as SDSS-RM project described by \citealt{Shen2015}) 
move forward, a large sample of AGNs will 
be available for searching for CB-SMBHs. A direct and efficient method is needed for preliminary 
selection of CB-SMBH candidates. In this supplementary paper, we calculate a series of atlases of 
2D TF for CB-SMBHs with various possible orbital and geometric parameters.
Given the continuum and velocity-resolved emission-line light curves from the RM campaign, 2D TFs 
of BLRs can be reconstructed straightforwardly using the maximum entropy method (MEM) 
\citep{Horne1994, Xiao2018a}.
Presenting complete features of 2D TFs for various CB-SMBH systems, the TFs 
extracted from current and future RM data can be directly compared against our
atlases to select CB-SMBH candidates efficiently. A few targets from the MAHA campaign show
promising signatures of CB-SMBHs (in a forthcoming paper). Even rough ranges of orbital parameters are 
possible to be estimated by direct comparison, which is important for further detailed analysis and confirmation.

\section{Reverberations of binary BLRs}
Reverberation mapping is a powerful tool to measure the kinematics of ionized 
gas around BHs.  The technique makes the following assumptions: (1) the 
origin of the ionizing photons (the accretion disk) is a point source much 
smaller than the BLR; (2) photoionization is the dominant ionization mechanism
for the emission lines; and (3) the BLR is relatively stable within the 
reverberation mapping timescale \citep[see reviews of][]{Peterson1993,
Peterson2014}. A single accreting BH is usually assumed as the default in the 
galactic center, but obviously this is inappropriate in the case of CB-SMBHs. 
Composition of reverberations of binary BLRs should be investigated for cases of CB-SMBHs, which are still spatially 
unresolved for direct imaging.

%
In the seminal paper of \citet{Blandford1982}, the reverberation of emission 
lines in response to continuum variations is mathematically described by the 
convolution
\begin{equation}
    L_{\ell}(v,t) = \int_{-\infty}^{\infty} \dd{t'} L_c(t') \Psi(v, t-t'),
\end{equation}
where $L_{\ell}(v,t)$ and $L_c(t')$ are the light curves of the emission lines 
and continuum, respectively, and $\Psi(v, t)$ is the 2D TF, which encodes the 
geometric and kinematic information of the BLRs. This function presents the 
echo strength of the emission line ($\ell$) in the space of the velocity of 
the line profile and delays. Observations provide data for the light curves of
the velocity bins and ionizing fluxes. From the convolution theorem, it follows
\begin{equation}
\label{eq:Psi_tot0}
    \Psi(v,t) =  \frac{1}{2\pi} \int_{-\infty}^{\infty} \dd{\omega} e^{i \omega t}  
    \left[ \frac{\tilde{L}_{\ell}(v,\omega)}{\tilde{L}_{c}(\omega)} \right] 
    \equiv \mathscr{F}^{-1} \left[ \frac{\tilde{L}_{\ell}(v,\omega)}{\tilde{L}_{c}(\omega)} \right], 
\end{equation}
where 
\begin{equation}
    \tilde{L}_{\ell,c}(\omega) = 
    \int_{-\infty}^{\infty} \dd{t} e^{-i \omega t} L_{\ell,c}(t) \equiv \mathscr{F}[L_{\ell,c}(t)],
\end{equation}
where $\omega$ is the angular frequency in Fourier space, and $(\mathscr{F},\mathscr{F}^{-1})$ 
are the Fourier transform and inverse Fourier transform, 
respectively.  We stress that this formal expression does not assume the case 
of a single SMBH and a single BLR.

The construction of 2D TFs of binary BLRs has been discussed by \citet{Wang2018}.
For convenience, we extend the derivation briefly here and use simulations to 
show the validity of the linear approximations. In this stage, we assume that 
the CB-SMBHs have their own BLRs, each ionized only by its own accretion disk. 
This assumption should be revised, as described in a forthcoming work, if the 
binary BLRs have merged to a later phase, in which case the CB-SMBH would be 
surrounded by a common BLR with an asymmetric geometry that is photoionized 
by two ionizing sources.  The two independently varying continuum sources 
are denoted as $L^{c}_{1,2}(t)$, and the corresponding broad emission lines 
are $L^{\ell}_{1,2}(v,t)$.  Then, for a spatially unresolved CB-SMBH, the 
total continuum and line emission are
\begin{equation}
\label{eq:flux_tot}
L_{\rm c}(t)=L_1^{\rm c}(t)+L_2^{\rm c}(t);\quad L_{\ell}(v,t)=L^{\ell}_1(v,t)+L^{\ell}_2(v,t).
\end{equation}
The linear combination of two independent components
can be expressed linearly in Fourier space:
\begin{equation}
\label{eq:Fourie_tot}
\tilde{L}_{\rm c}(\omega)=\tilde{L}^{c}_{1}(\omega) + \tilde{L}^{c}_{2}(\omega);\quad
\tilde{L}_{\ell}(v,\omega)=\tilde{L}^{\ell}_{1}(v,\omega) + \tilde{L}^{\ell}_{2}(v,\omega).
\end{equation}
Inserting Equation (\ref{eq:Fourie_tot}) into (\ref{eq:Psi_tot0}), the 2D TF is
\begin{equation}
	\Psi_{\rm tot}(v,t) 
	= \mathscr{F}^{-1} \left[ \frac{\tilde{L}^{\ell}_{1}(v,\omega) + 
	\tilde{L}^{\ell}_{2}(v,\omega)}{\tilde{L}^{c}_{1}(\omega) + \tilde{L}^{c}_{2}(\omega)} \right]
	=\mathscr{F}^{-1} \left[\frac{\tilde{L}^{\ell}_{1}(v,\omega)}{\tilde{L}^{\rm c}_{1}(\omega)+
	 \tilde{L}^{\rm c}_{2}(\omega)}+  
	 \frac{\tilde{L}^{\ell}_{2}(v,\omega)}{\tilde{L}^{\rm c}_{1}(\omega)+
	 \tilde{L}^{\rm c}_{2}(\omega)}\right].
\end{equation}	
Introducing the individual 2D TFs, we have 
\begin{equation}	
\label{eq:Psi_tot}
\Psi_{\rm tot}(v,t)	= \mathscr{F}^{-1} \left[ \frac{\calL_1(v,\omega)}{1 + \Gamma_{\omega}} 
                     + \frac{\calL_2(v,\omega)}{1 + \Gamma_{\omega}^{-1}} \right]
                    =\mathscr{F}^{-1}\left[ \frac{\calL_1(v,\omega)}{1 + \Gamma_{\omega}}\right] 
                     +\mathscr{F}^{-1}\left[\frac{\calL_2(v,\omega)}{1 + \Gamma_{\omega}^{-1}} \right],
\end{equation}
where 
$$\calL_{1,2}(v,\omega) \equiv \tilde{L}^{\ell}_{1,2}(v,\omega)/\tilde{L}^{c}_{1,2}(\omega),$$ and 
the coupling coefficient is given by
%
$\Gomega \equiv \tilde{L}^{c}_{2}(\omega)/\tilde{L}^{c}_{1}(\omega)$,
%
which is caused by the limitation of spatial resolution of the telescope.
The quantity $\Gomega$ cannot be given analytically a priori, but we 
can derive it from the 
properties of the optical variability of AGNs with single BHs.
This paper provides more discussion on this coupling coefficient.

It should be noted that $\Psi_{1,2}(v,t) = \mathscr{F}^{-1} [ \calL_{1,2}(v,\omega)]$, 
where $\Psi_{1,2}(v,t) $ are the single TFs of the two BLRs.
Using the convolution theorem again, we have
\begin{equation}
	\Psi_{\rm tot}(v,t) = \sum_{k=1}^{2} \int_{-\infty}^{\infty} \Psi_{k}(v,t') \calQ_k(t-t') \dd{t'},
\end{equation}
where 
\begin{equation}
\calQ_1(t) \equiv \mathscr{F}^{-1}\left(\frac{1}{1+\Gomega}\right),\quad 
\calQ_2(t) \equiv \mathscr{F}^{-1}\left(\frac{1}{1+\Gomega^{-1}}\right).
\end{equation}
The sum of $\calQ_1(t)$ and $\calQ_2(t)$ is
\begin{equation}\label{eq:sum}
\calQ_1(t) + \calQ_2(t) = \mathscr{F}^{-1}\left(\frac{1}{1+\Gomega} + \frac{1}{1+\Gomega^{-1}}\right) = \mathscr{F}^{-1}(1) = \delta(t).
\end{equation}

If $\Gamma_{\omega}$ is a real constant $\Gamma_0$, 
we find that $\calQ_1(t) = \delta(t)/(1+\Gamma_0)$ and $\calQ_2(t) = \delta(t)/(1+\Gamma_0^{-1})$,
and so
\begin{equation}
\Psi_{\rm tot}(v,t) = \frac{\Psi_{1}(v,t)}{1+\Gamma_0} +  \frac{\Psi_{2}(v,t)}{1+\Gamma_0^{-1}}.
\end{equation}
That is, the total TF is a linear combination of two individual TFs.
However, $\Gamma_{\omega}$ generally is not a constant.
Given each BLR and the coupling coefficient, we can specify the composite 2D TF.

\section{The coupling coefficient}
\subsection{Generating light curves}
A number of monitoring campaigns have shown that AGN brightness varies as a continuous 
stochastic process, which can be described by the power spectral density (PSD) of its variation.
Earlier optical observations find that the PSD of quasars is a single power law with index $\gamma\approx 2$ \citep{Giveon1999}, consistent with a damped random walk (DRW, \citealt{Li2013, Zu2013}).  However, there is growing evidence for deviations from the DRW model, with $\gamma=2.13_{0.06}^{+0.22}$ in 13 AGNs \citep{Collier2001}, $\gamma=1.77$ in MACHOS quasars \citep{Hawkins2007}, and $\gamma=1.75-3.2$ in {\it Kepler} quasars \citep{Smith2018}. For the present discussion, we generalise the PSD in the form 
%
%
\begin{equation}\label{psf}
{\rm PSD}(f) = \frac{2 c_{\gamma} \tau_0 \sigma^2 }{1+(2\pi f \tau_0)^{\gamma}},
\end{equation} 
where $\tau_0$ is the characteristic timescale of variations, $\sigma$ is its 
amplitude, $\gamma$ is the power-law index, and $c_{\gamma}$ is a normalization
constant.  There are two regimes of interest: (1) when $f\ll \tau_0^{-1}$, 
${\rm PSD}$ is a constant; (2) when $f\gg\tau_0^{-1}$, ${\rm PSD}\propto 
f^{-\gamma}$.  The present PSD functions fulfill the conditions of most 
observations, and hence satisfy the requirements for $\Gomega$.

As shown by \cite{Timmer1995}, the PSD of variations is the Fourier transform 
of its auto-covariance function (ACF).  That is,
\begin{equation}
{\rm ACF} (\tau) \equiv \langle \Delta M(t) \Delta M(t+\tau) \rangle = \int_{-\infty}^{\infty} \dd{f} e^{i2\pi f\tau} {\rm PSD}(f),
\end{equation}
where $\Delta M(t) = M(t) - \langle M \rangle$ is the deviation of the AGN's 
magnitude from its expectation value at time $t$. In practice, $c_{\gamma}$ can be obtained from
\begin{equation}
2\int_{0}^{\infty} \dd{f} {\rm PSD}(f) = \sigma^2.
\end{equation}
As a natural consequence, we have $\langle (\Delta M(t))^2 \rangle = \sigma^2$ in our formulation.  Given the PSD of the variations, realization of the process 
(i.e. light curve of the AGN) can be simulated by the algorithm, which is easy 
to implement \citep{Timmer1995}. 
The major steps are as follows:
(1) For each angular frequency $\omega$, draw a random number from  a  chi-squared distribution with 2 degrees of freedom, then multiply it by ${\rm PSD}(f = \omega/2\pi)/2$.  This will be used as the module of the variation in Fourier space $\Delta \tilde{M}(\omega)$.
(2) The phase angle of $\Delta \tilde{M}(\omega)$  is generated uniformly between $0$ and $2\pi$.
(3) The inverse Fourier transform of $\Delta \tilde{M}(\omega)$ yields $\Delta M(t)$, and $M(t)$ is given by $\langle M \rangle + \Delta M(t)$.
(4) The magnitude $M(t)$ now can be converted directly to luminosity $L(t)$. 
%
This scheme allows us to test the properties of the coupling coefficient ($\Gomega$).

\subsection{$\calQ-$dependence}
\begin{figure}
\plotone{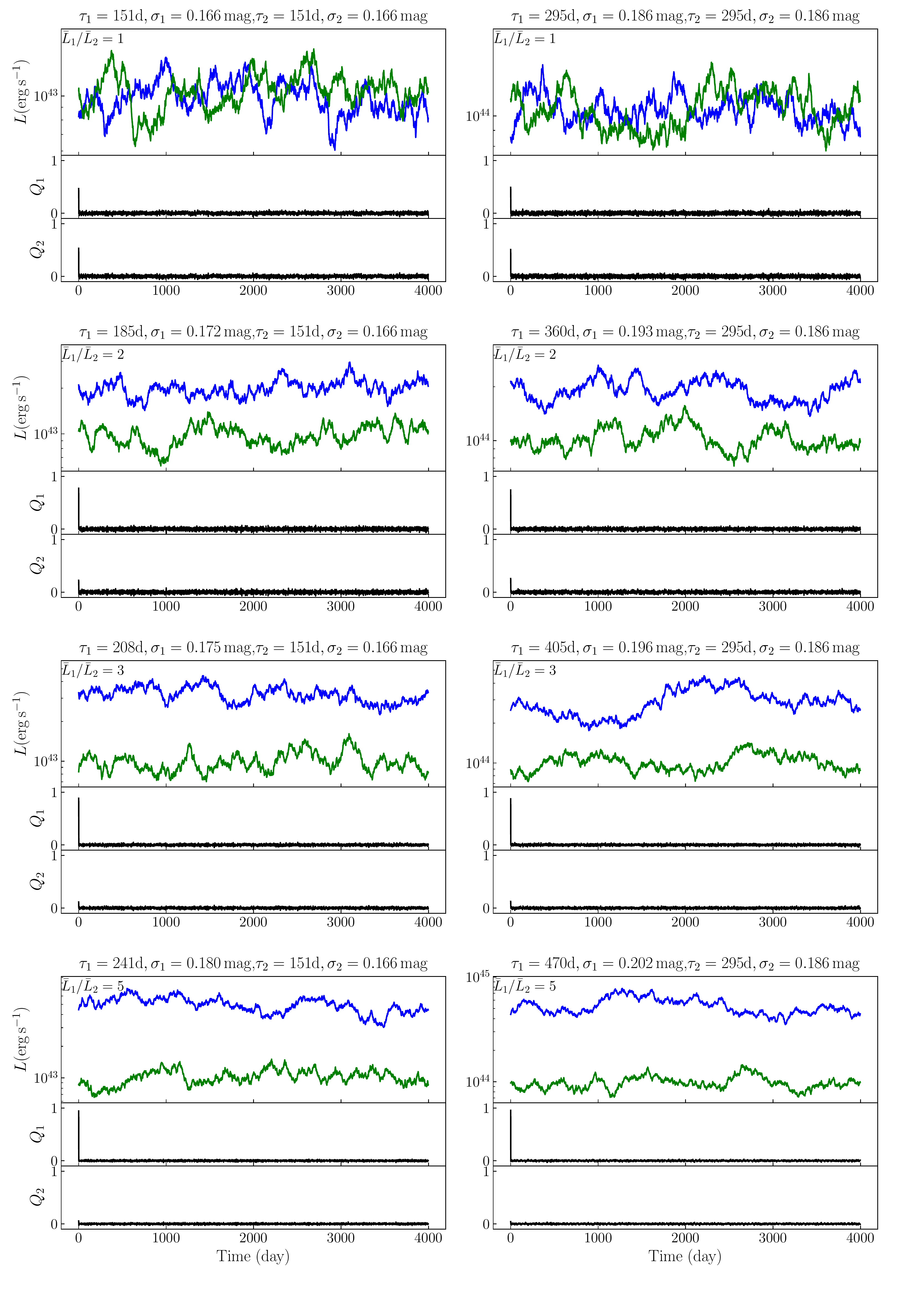}
\caption{\footnotesize \label{fig:f1} Examples of simulated continuum light curves and corresponding $\calQ$ functions.
In the left panels, the average $L_{5100}$ of the secondary BHs are all $10^{43} \, {\rm erg \, s^{-1}}$, and the luminosity ratios are $1$, $1/2$, $1/3$ and $1/5$ from top to bottom.
In the right panels, the average $L_{5100}$ of the secondary BHs are all $10^{44} \,{\rm erg \, s^{-1}}$, and the luminosity ratios are also $1$, $1/2$, $1/3$ and $1/5$ from top to bottom.
The timescales and amplitudes of variations are obtained through equation (\ref{kelly}).}
\end{figure}

In order to test the $\calQ-$dependence, we use correlations of $\tau_0$ and $\sigma$ with optical luminosity from \citet{Kelly2009}, who analyzed a sample of AGN optical light curves and modeled them as 
a stochastic process with $\gamma = 2$. Given the 5100 \AA\, luminosity, 
$\tau_0$ and $\sigma$ are found from the intermediate correlations 
with AGN luminosity, which are given by
\begin{equation}\label{kelly}
\log \tau_0 = -10.29 \pm 2.76 + (0.29 \pm 0.08) \log L_{5100},\quad 
\log \sigma^2 = -(5.86 \pm 4.43) + (0.10 \pm 0.09) \log L_{5100}.
\end{equation}
%
Here the $\sigma$ correlation is converted from \cite{Kelly2009}, as their definition of $\sigma$ is different from ours.
Although the above relations assume $\gamma=2$, we will utilize them in our simulations to generate light curves for $\calQ$.  
Examples of simulated continuum light curves and the corresponding $\calQ_{1,2}(t)$ are illustrated in Figure \ref{fig:f1}.

\begin{figure}
\plotone{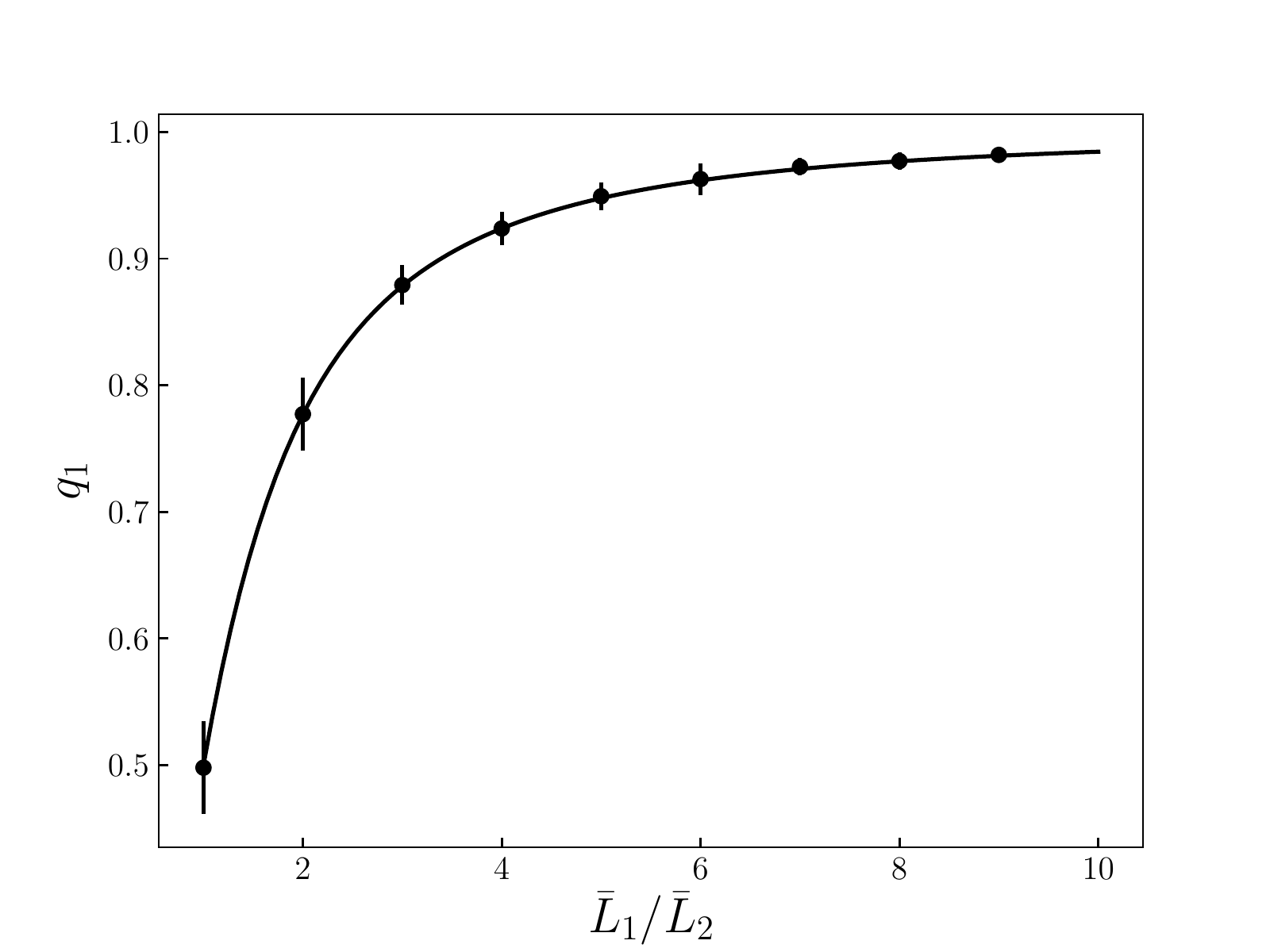}
\caption{\footnotesize \label{fig:fr} The correlation between the coefficient of linear combinations and luminosity ratios. 
The average $L_{5100}$ of the secondary BH is taken to be $10^{44} \,{\rm erg \, s^{-1}}$.
The timescales and amplitudes of variations are obtained through equation (\ref{kelly}).
For every data point, $100$ light curves are generated to calculate the value and uncertainty for the coefficient $q_1$.
We also note that the correlation can be well described by an empirical formula, $q_1 = 1 - 1 / [(\bar{L}_1/\bar{L}_2)^{1.8}+1]$ (solid line).
}
\end{figure}

For a wide range of average luminosity and luminosity ratios, our simulation shows that $\calQ_{1}(t)$ for the primary BH can always be treated as delta
functions with minor noise. 
If we assume $\calQ_{1}(t) \approx q_1\delta(t)$, we immediately will obtain $\calQ_2(t) \approx (1-q_1)\delta(x)$ from Equation (\ref{eq:sum}).
Thus, the total transfer function can be calculated as
\begin{equation}\label{tot}
\Psi_{\rm tot}(v,t) = q_1 \Psi_1(v,t) + (1-q_1) \Psi_2(v,t).
\end{equation}
The coefficient $q_1$ is correlated with the luminosity ratio in our simulations, as shown in Figure \ref{fig:fr}.

Nevertheless, we should mention that the noise amplitude of $\calQ_{1,2}(t)$ depends on the cadence and duration of the sampling. Only if the cadence is much shorter than the timescale of the variation of both continuum sources while the duration is much longer will our linear approximation be valid.
This is reasonable since it is almost impossible to extract TF from poor RM data with bad cadence 
and short duration, even in the case of a single BLR.


\section{BLR Models and Transfer Functions}
\subsection{Kinematics}
Much progress has been made in reverberation mapping campaigns during the last 
three decades, particularly in the application of the velocity-resolved 
technique to study the kinematics of the BLR.  There is growing evidence that 
the BLRs in most AGNs have rather simple geometry and dynamics. Flattened disks
are common in Seyfert galaxies \citep{Grier2013}, even among narrow-line
Seyfert 1 galaxies \citep{Du2016}. Inflows or outflows have been reported in a 
few cases. Detailed investigations of a few objects (e.g., NGC 5548, 3C 390.3 
and NGC 7469; \citealt{Wandel1999,Lu2016}) show that the full width at  
half-maximum (FWHM) of H$\beta$ and its lag follows 
$\tau_{\rm H\beta}\propto {\rm FWHM}^{-1/2}$, indicating a Keplerian rotating
disk.  These conclusion is supported by detailed dynamical modeling
\citep{Pancoast2011, Pancoast2014a, Pancoast2014b, Grier2017,Li2013, Li2018b}. 
This paper only focuses on models in which the binary BLRs are still 
independent so that each BLR can be described by a flattened disks. We do not 
exclude the presence of inflows or outflows, but here we 
are mainly concerned with the disk-like geometry of the BLRs.  

Table \ref{tab:par} lists the parameters of the present model. For each BLR, 
the main parameters are the inner and outer radii ($R_{\rm in}^{i}$ and 
$R_{\rm out}^{i}$), the power-law index of the reprocessing coefficient 
($\gamma_{i}$ for the two BLRs), and the half-opening angle ($\Theta_{i}$). 
The binary BLRs are rotating around the center of mass of the binary BHs,
and the binary disks are aligned with the orbital plane. Each flattened BLR is 
allowed to rotate in the same or completely opposite direction as the orbital 
motion. In future, polarised spectra or DPCs obtained by GRAVITY 
\citep{Songsheng2019} may be able to resolve the sense of rotation, but this 
effect cannot be distinguished in the total spectra.

Given the velocity field and the reprocessing coefficient distribution of each 
BLR, the 2D TF in principle can be calculated according to 
\cite{Blandford1982}.  Suppose that the velocity distribution of the clouds in 
BLR $i$ at a given point is $f_i(\bm{r}_i,\bm{V}_i)$, where $\bm{r}_i$ is the 
displacement to its central BH and $\bm{V}_i$ is the velocity of the cloud.
The reprocessing coefficient at that point is $\xi(\bm{r})$.
The 2D TF for the BLR is therefore
\begin{equation}
\Psi_i(v,t) = \int \dd{\bm{r}_i} \dd{\bm{V}_i} \frac{\xi_i(\bm{r}_i) f_i(\bm{r}_i,\bm{V}_i)}{4\pi r_i^2}
\delta(v - \bm{V}_i\vdot\bm{n}_{\rm obs}) \delta(ct - r_i - \bm{r}_i \vdot \bm{n}_{\rm obs}),
\end{equation} 
where $\bm{n}_{\rm obs}$ is the unit vector pointing from the observer to the source.
However, the orbital motion should be included in the velocity field of each individual BLR.
It should be noted that the composite velocities of clouds is the superposition of the individual virial motion (or inflow/outflow motion) and orbital motion of the binary system.  Thus,
$\bm{V}_i = \bm{V}_{\rm vir} + \bm{\Omega} \times (\bm{A}_i + \bm{r}_i)$, 
where $\bm{\Omega}$ is the angular velocity of the system, $\bm{A}_i$ is the displacement of the central BH from the center of mass of the system, and $\bm{V}_{\rm vir}$ is the virial velocity of the cloud in the co-rotating frame. 
As the rotation period of the system (tens or hundreds of years) is much longer than the time scale of reverberation (tens or hundreds of days), 
$f_i(\bm{r}_i,\bm{V}_i)$, $\xi_i(\bm{r}_i)$ and $\bm{A}_i$ are considered to be invariant when calculating the TFs.


\subsection{Atlas of 2D TFs}
\begin{figure}
\plotone{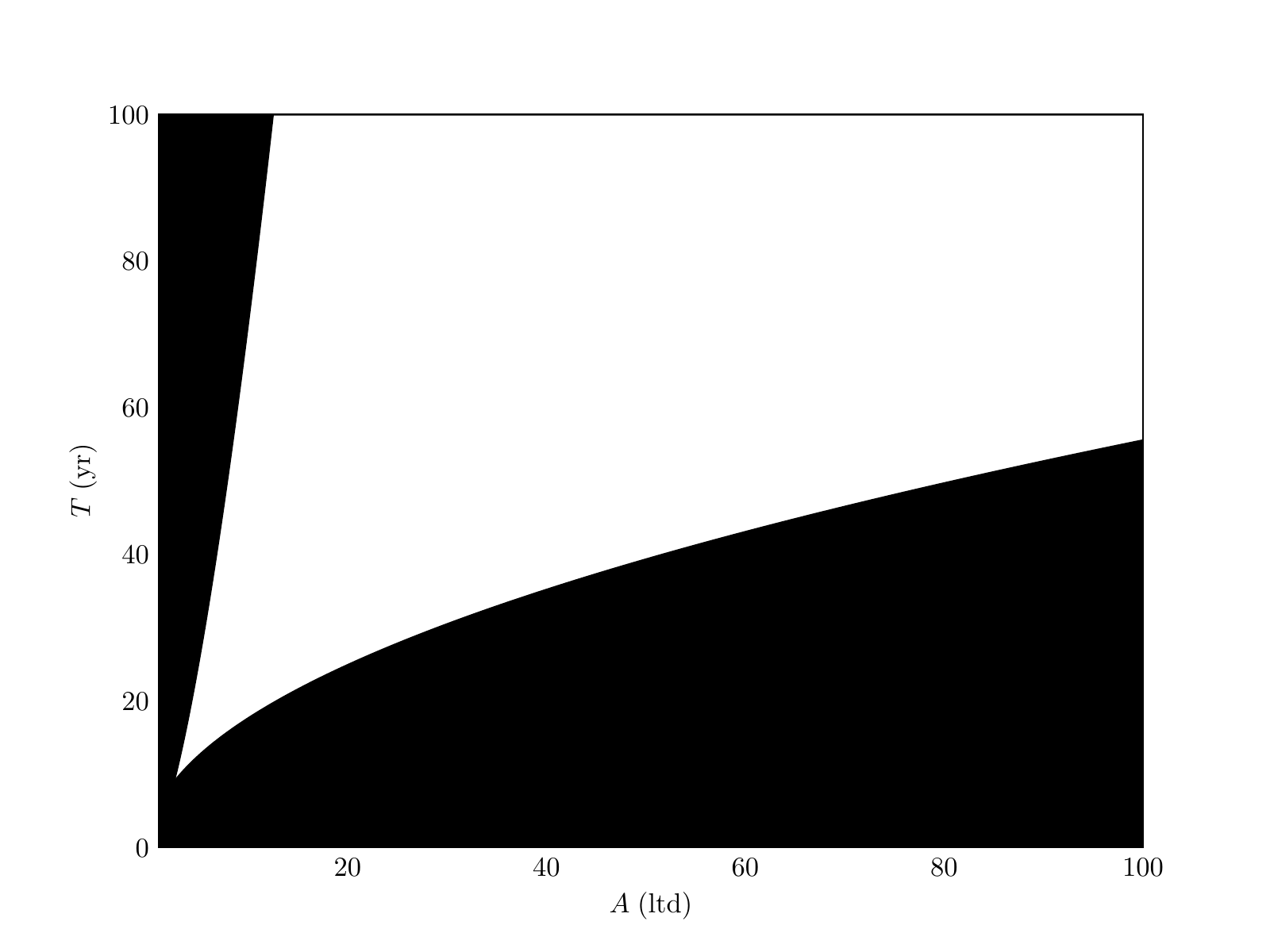}
\caption{\footnotesize 
\label{fig:f2} Allowed region in the $A-T$ plane when $\mu_1 = 0.7$. Only parameters in the white region are 
valid. Ranges of allowed regions for different $\mu_1$ are very similar. The rotation period for binary 
BLRs before merging is mostly larger than $20$ yr.
}
\end{figure}

This paper focuses on the separated binary BLRs of CB-SMBHs. The individual BLR is assumed to follow the scaling relation between size and optical luminosity as $R_{\rm BLR}\approx 33.6\,L_{44}^{0.533}$ltd, where $L_{44} = L_{5100}/10^{44}\,{\rm erg\,s^{-1}}$\citep{Bentz2013}.
It should noted that this only applies to sub-Eddington AGNs \citep{Du2014,Du2015,Du2018a}.
Given the BH mass, the optical luminosity at 5100 \AA\, can be calculated by
\begin{equation}
L_{5100} = 1.26 \times 10^{43} \lambda_{0.01} \kappa_{10}^{-1} M_{8} \; {\rm erg \, s^{-1}},
\end{equation}
where $\lambda_{0.01} = \lambda_{\rm Edd}/0.01$ is the Eddington ratio and 
$\kappa_{10} = \kappa_{\rm bol}/10$ is the bolometric correction factor. The average radius of 
the BLR can be expressed as 
\begin{equation}
\langle R_{\rm BLR} \rangle \approx 11.2 \lambda_{0.01}^{1/2}\kappa_{10}^{-1/2}M_{8}^{1/2} \, {\rm ltd}.
\end{equation}
%
The average radius of BLR is defined by
\begin{equation}
\langle R_{\rm BLR} \rangle =\frac{\int_{R_{\rm in}}^{R_{\rm out}}r \xi(r) \dd{r}}
                              {\int_{R_{\rm in}}^{R_{\rm out}}\xi(r)\dd{r}},
\end{equation}
where $\xi(r) \propto (r/R_{\rm in})^{-\gamma}$ is the reprocessing coefficient.
Given the separation between the two BHs and the rotation period of the binary system, the Keplerian relation
can be expressed by Equation (\ref{eq:period}).
We choose the BH mass from $10^6$ to $10^{10}\,M_{\sun}$. 
Since we assume that the CB-SMBHs have their own BLRs and each ionized only by its own accretion disk, the two BLRs must be well separated.
Thus, there is another additional constraint on the separation,
\begin{equation}
R_{\rm out,1} + R_{\rm out,2} < A,
\end{equation}
which limits the space of the CB-SMBH parameters. In our calculations, we will assume $\lambda_{0.01} = 1$ and $\kappa_{10}=1$ for both BLRs. The inner and outer radii of the BLRs will then be taken as $0.52 \langle R_{\rm BLR} \rangle$ and $1.57\langle R_{\rm BLR} \rangle$, corresponding to $\gamma=0.5$ and $R_{\rm out}/R_{\rm in} = 3$. The masses of the primary and secondary BH can be obtained through $M_1 = \mu_1 M_{\rm tot}$ and $M_2 = (1-\mu_1)M_{\rm tot}$, where $\mu_1$ is the mass ratio with respect to the total mass. 
Figure \ref{fig:f2} shows the allowed region in the $A-T$ plane.  In this paper, we only consider the cases with $\mu_1\le 0.9$.  Systems with $\mu_1\le 0.9$
may have very different kinematics, and will be treated in a forthcoming paper.

The opening angle of the flattened Keplerian disk and the inflows/outflows will be fixed at $10^{\circ}$ and $45^{\circ}$, respectively.
The power-law index of the gas distribution is taken to be $\gamma = 0.5$.
In the case of inflows, the velocity of the gas at $\bm{R}$ is assumed to be $-1.4V_{\rm K}\bm{e}_r$, which is slightly smaller than the local escape velocity.
For outflows, the velocity is $1.6(r/R_{\rm out})^{0.1} V_{\rm K} \bm{e}_r$, slightly higher than the local escape velocity, such that 
the gas gains energy steadily when escaping outward.

%
In simulations of continuum light curves and $\calQ_{1,2}(t)$ shown in Figure~\ref{fig:f1}, we find that the coefficient $\Gamma_0$ in linear combinations of individual TFs is generally correlated with the ratio of the average optical luminosity of the two BHs.
We also assume that the optical luminosity is proportional to the mass of the BH. 
As a crude simplification, the total TF will be obtained from the individual TFs through
\begin{equation}
\Psi_{\rm tot} = \mu_1 \Psi_{1} + (1-\mu_1) \Psi_{2}.
\end{equation}

\noindent
We have five free parameters to vary when making the atlas of 2D TFs for binary BLRs.  The separation $A$ and period $T$ must lie within the white area of 
Figure~\ref{fig:f2}: we take $(A,T) = (10\, {\rm ltd}, 20\, {\rm yr})$, $(10\, {\rm ltd}, 50\, {\rm yr})$, $(20\, {\rm ltd}, 50\, {\rm yr})$, $(20\, {\rm ltd}, 100\, {\rm yr})$, $(50\, {\rm ltd}, 50\, {\rm yr})$, $(50\, {\rm ltd}, 100\, {\rm yr})$, and $(100\, {\rm ltd}, 100\, {\rm yr})$.  For each pair of $(A,T)$, the mass fraction of the primary BH will take values of $\mu_1 = 0.6$, $0.7$, and $0.8$, the inclination angle along the LOS will pass through $i = 15^{\circ}$, $30^{\circ}$, and $45^{\circ}$, and the phase angle of the rotation will assume values of $\phi = 0^{\circ}$, $45^{\circ}$, $90^{\circ}$, $135^{\circ}$, and $180^{\circ}$ to cover half a rotation period.

The atlases of 2D TFs are shown in the following pages.  The TF of a binary BLR consisting of two thin disks is composed of two shift bells.  The height of each bell is mainly determined by the size of the BLR, while the width of the bell reflects the maximum velocity of clouds along the LOS, determined by the mass of its BH, the size of BLR and inclination of LOS.  The separation between the two bells is governed by the orbital velocity of the binary BHs along the LOS, determined by the distance between two BHs, orbital period, orbital phase, and inclination.

\begin{figure}
\plotone{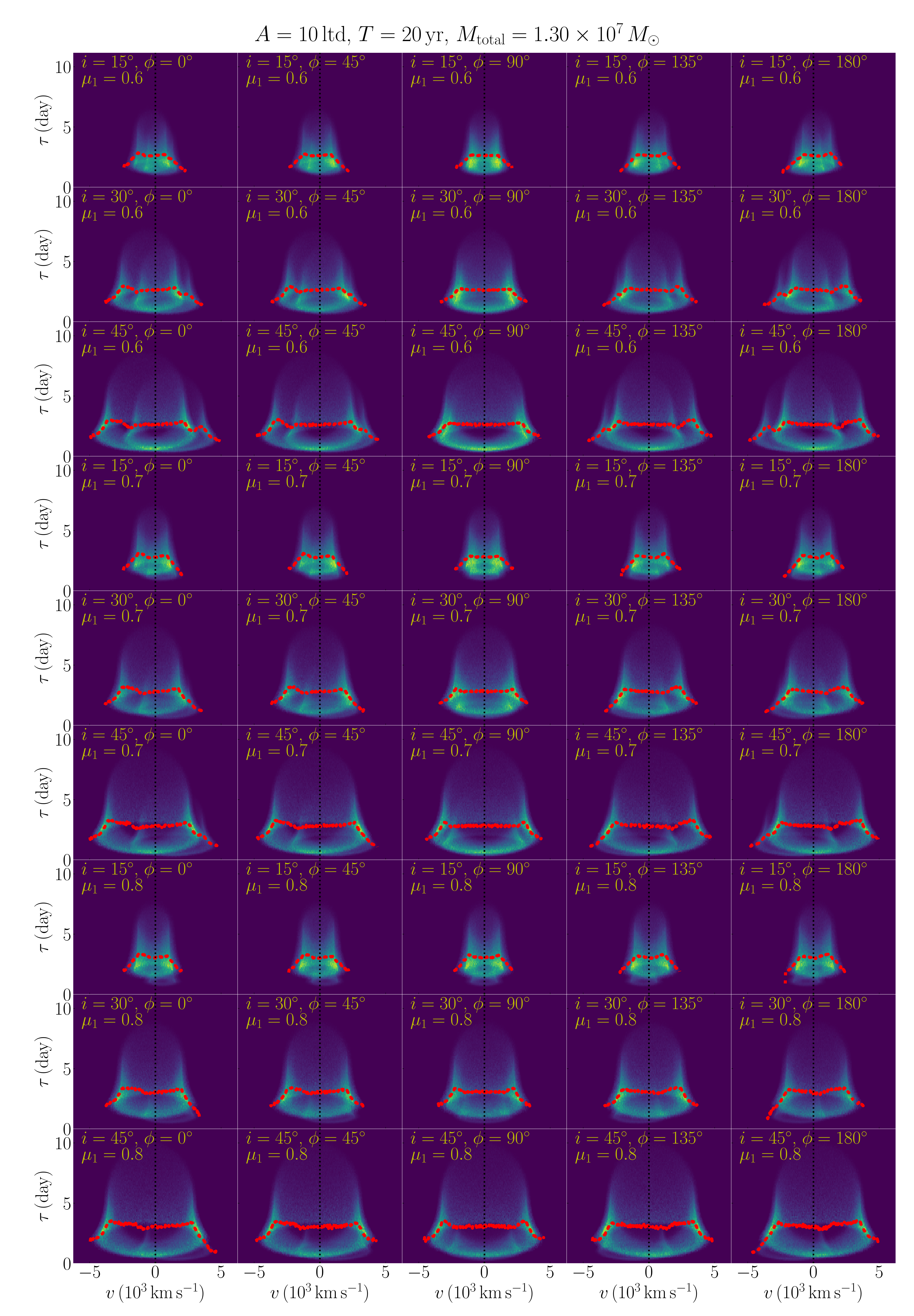}
\caption{\label{fig:trans_3_4} Atlas of 2D TFs of two disk-like BLRs with $A = 10$ ltd and $T = 20$ yr.}
\end{figure}

\begin{figure}
\plotone{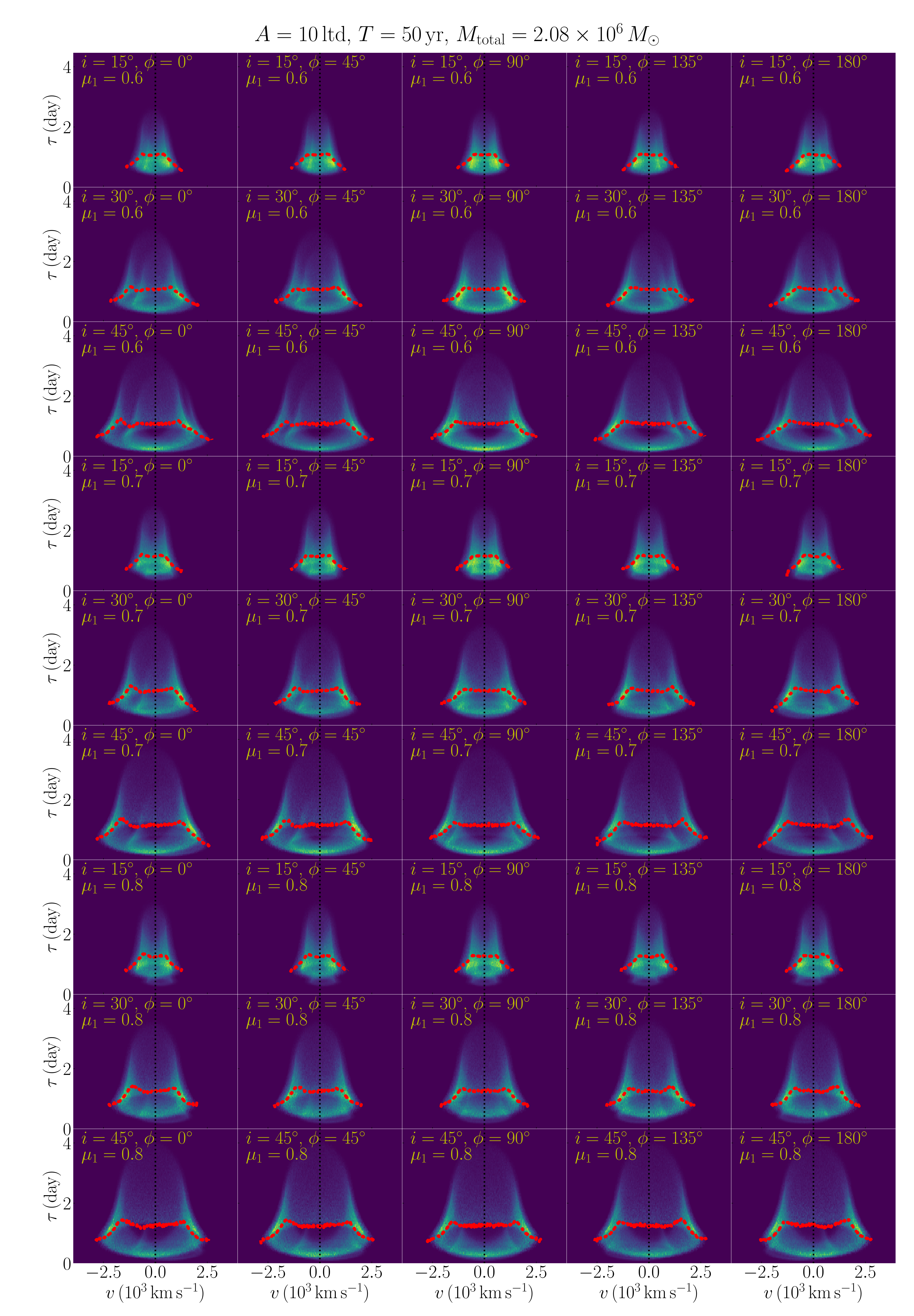}
\caption{\label{fig:trans_3_5} Atlas of 2D TFs of two disk-like BLRs with $A = 10$ ltd and $T = 50$ yr.}
\end{figure}

\begin{figure}
\plotone{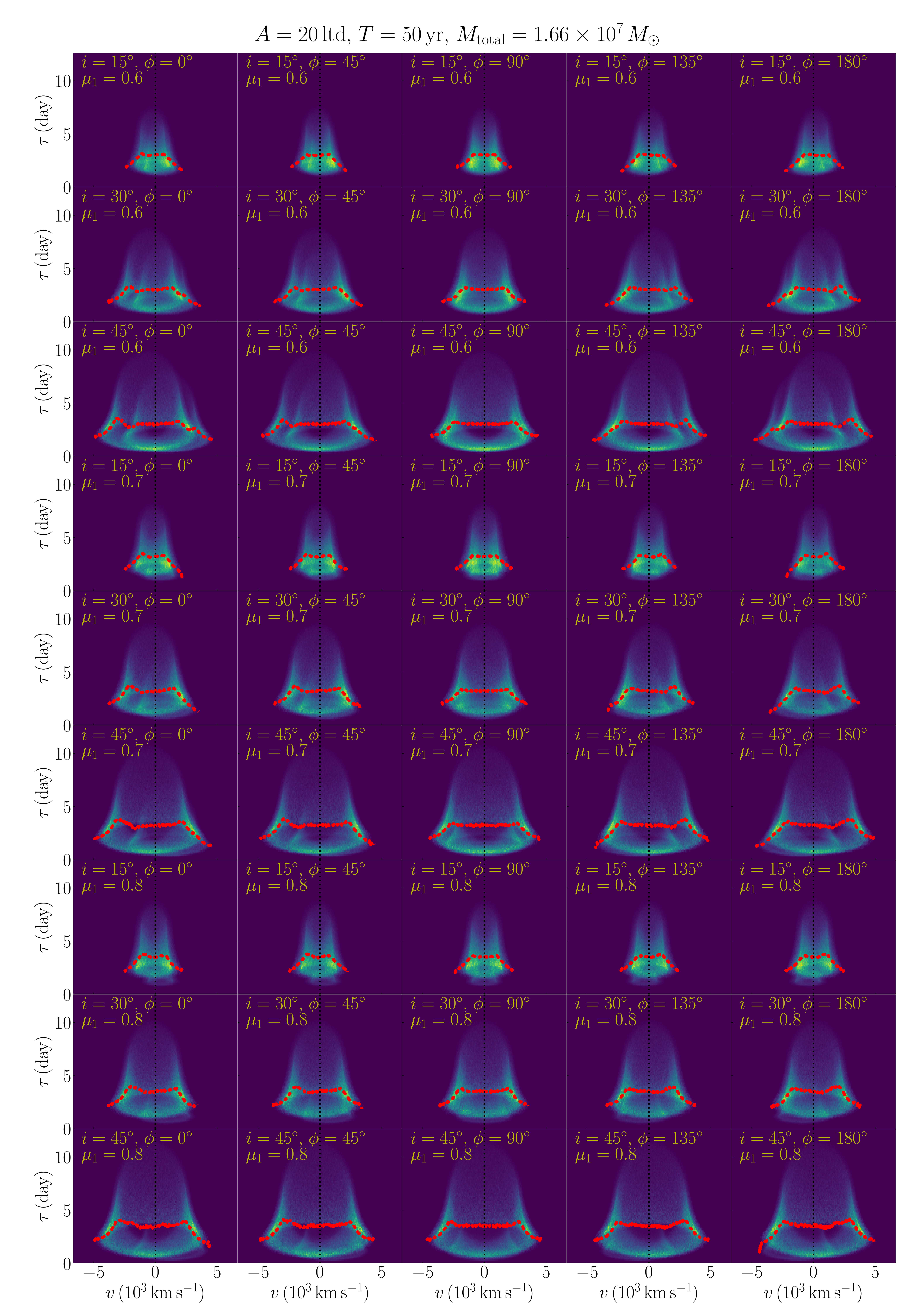}
\caption{\label{fig:trans_4_5} Atlas of 2D TFs of two disk-like BLRs with $A = 20$ ltd and $T = 50$ yr.}
\end{figure}

\begin{figure}
\plotone{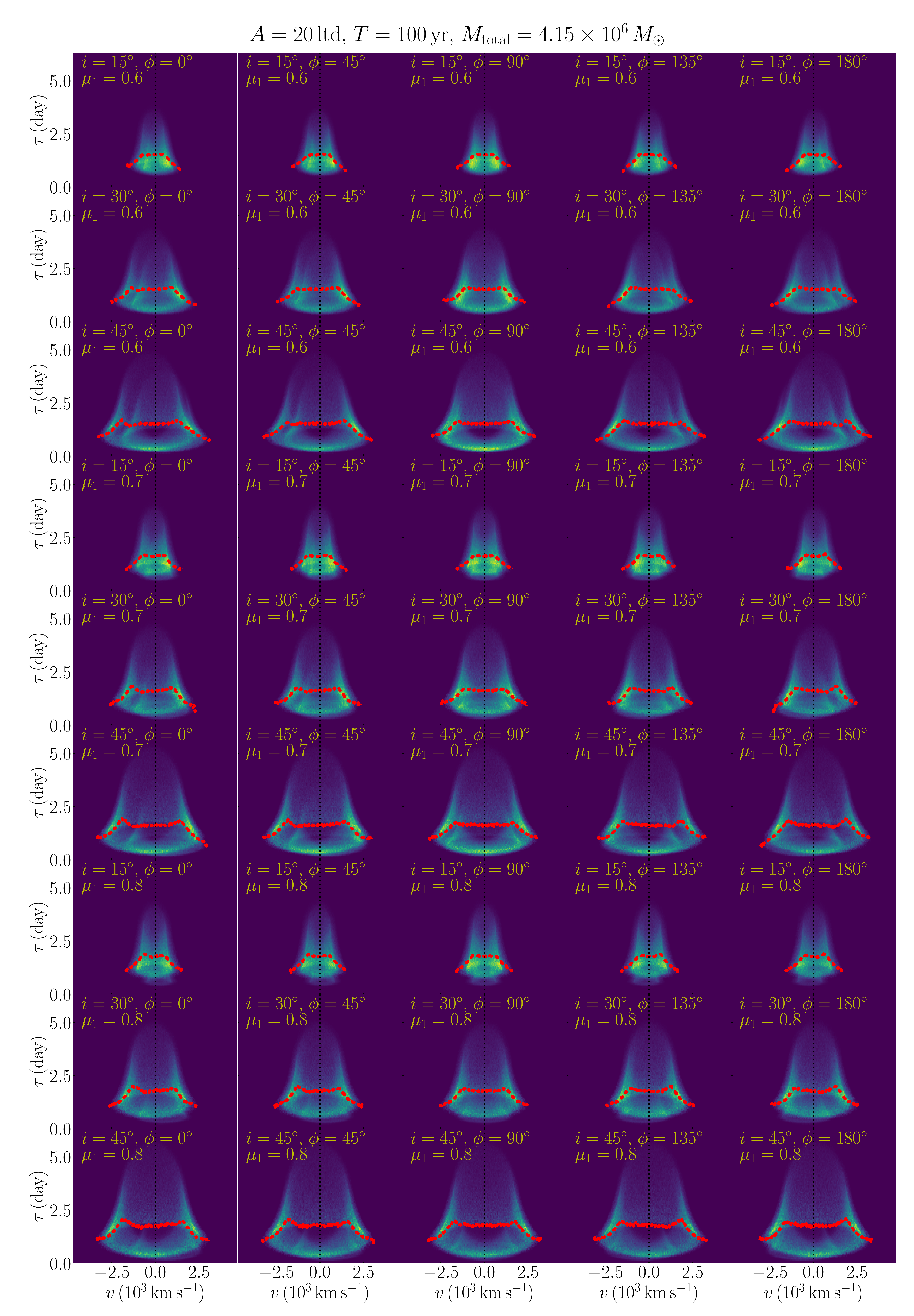}
\caption{\label{fig:trans_4_6} Atlas of 2D TFs of two disk-like BLRs with $A = 20$ ltd and $T = 100$ yr.}
\end{figure}

\begin{figure}
\plotone{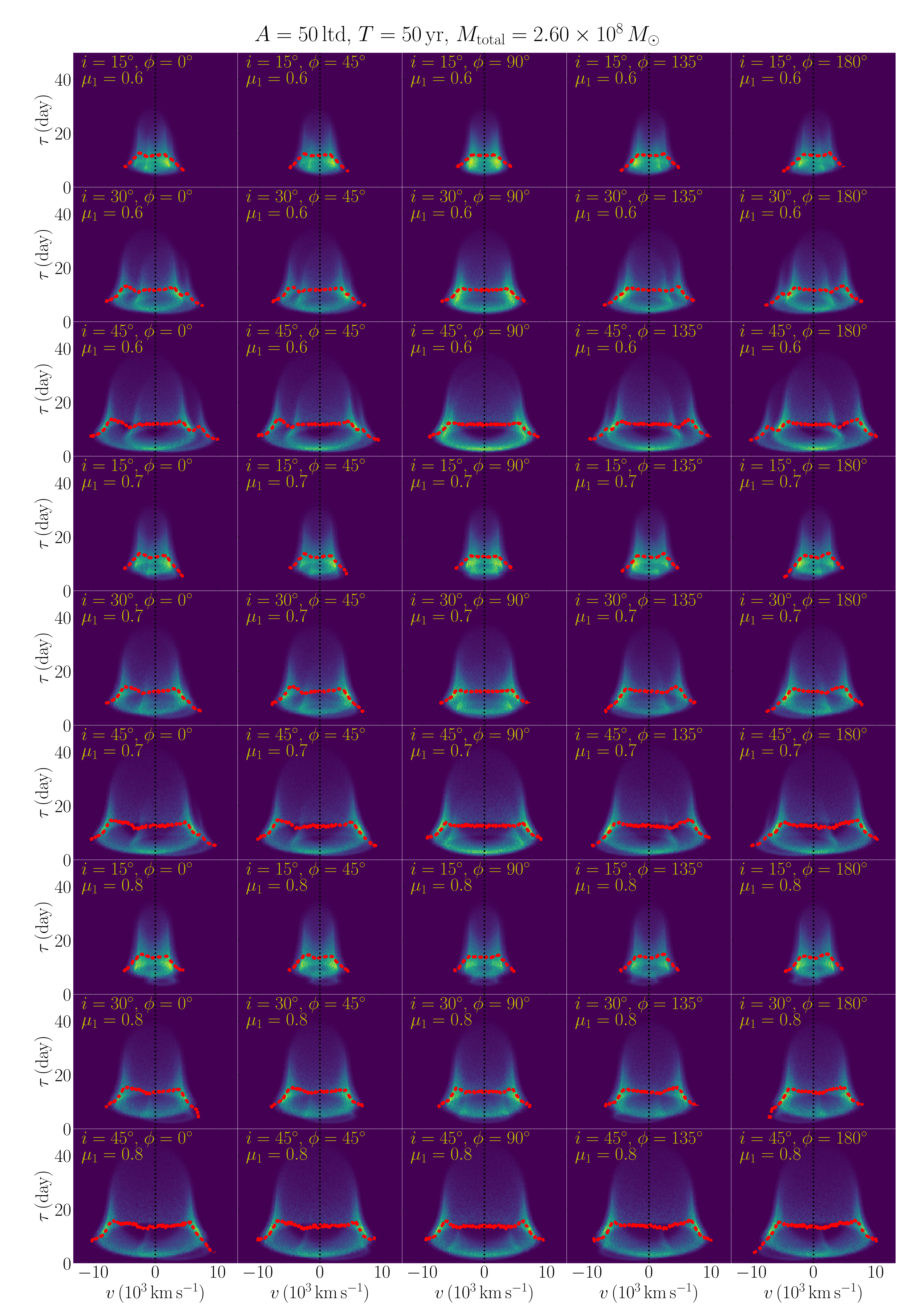}
\caption{\label{fig:trans_5_5} Atlas of 2D TFs of two disk-like BLRs with $A = 50$ ltd and $T = 50$ yr.}
\end{figure}

\begin{figure}
\plotone{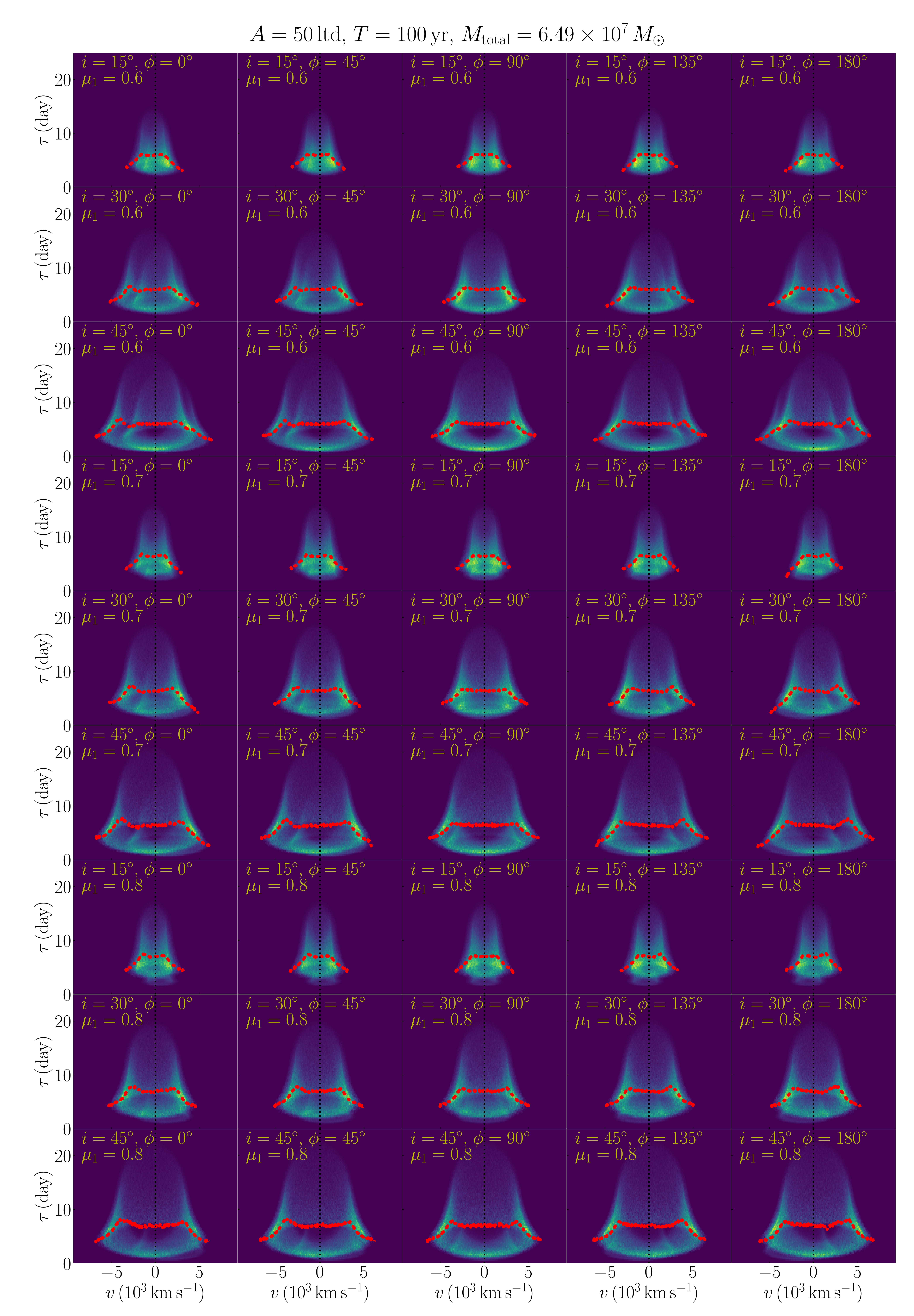}
\caption{\label{fig:trans_5_6} Atlas of 2D TFs of two disk-like BLRs with $A = 50$ ltd and $T = 100$ yr.}
\end{figure}

\begin{figure}
\plotone{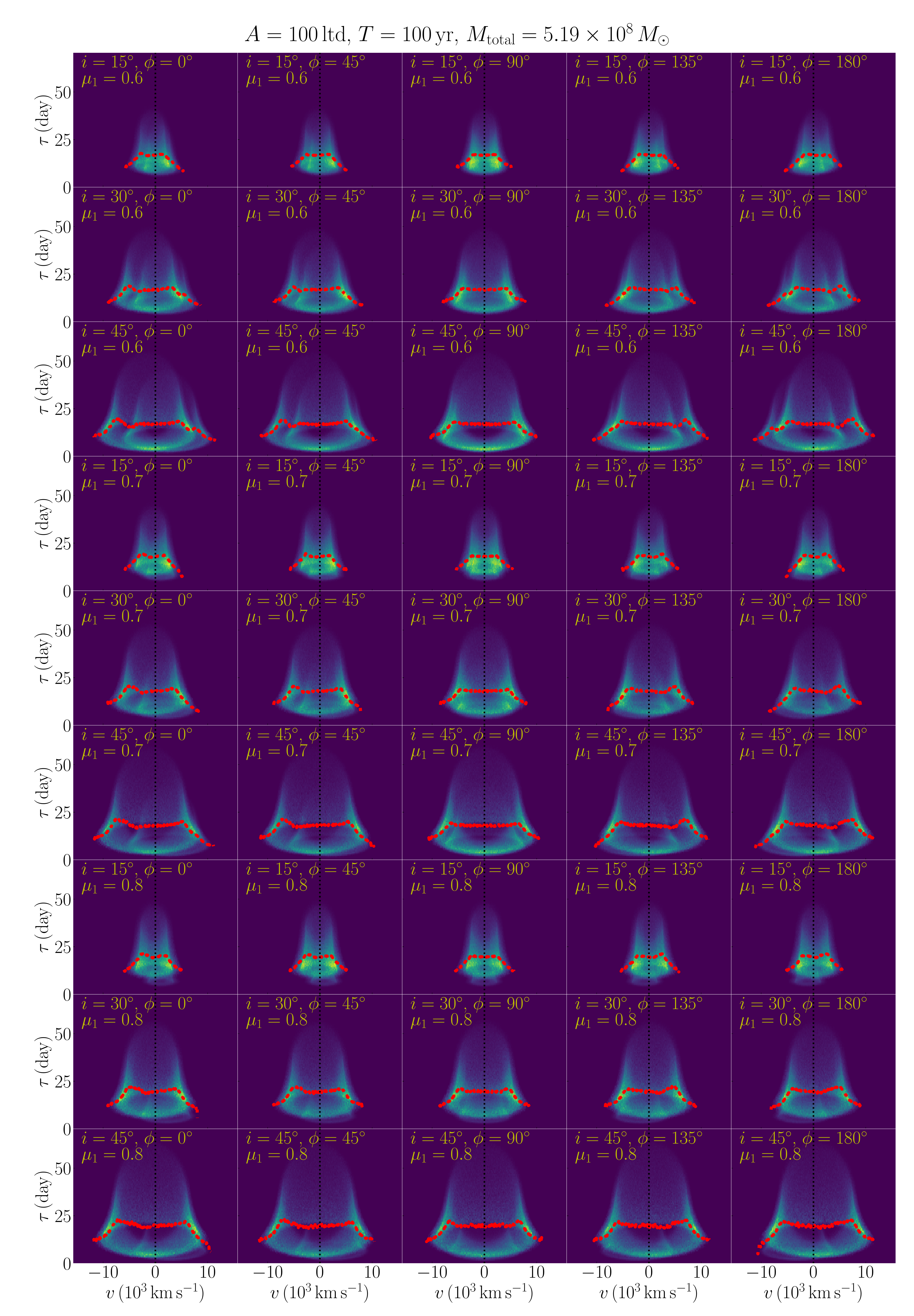}
\caption{\label{fig:trans_6_6} Atlas of 2D TFs of two disk-like BLRs with $A = 100$ ltd and $T = 100$ yr.}
\end{figure}

From Figure \ref{fig:trans_3_4}, it can be seen that when the phase angle increases from $0^{\circ}$ to $180^{\circ}$ the two bells of the TF approach each other in the beginning, and then separate and move to the opposite side.  When the inclination increases, the width and separation of the bells will become larger owing to the increase of projected velocity.  The edge of the bell also becomes sharper when the BLRs are viewed edge-on.  The mass ratio of the primary BH adjusts the relative size of the two bell shapes in the TF.  When $\mu_1 \gtrsim 0.8$, the effect of the secondary BLR will be hard to detect if we assume that the Eddington ratios of two BHs are comparable.
The 2D TFs of two disk-like BLRs with other combinations of ($A$, $T$) are shown in Figures \ref{fig:trans_3_5}--\ref{fig:trans_6_6}.  Their shapes and dependence on inclination, orbital phase, and mass ratio are similar to the case in Figure~\ref{fig:trans_3_4}. Only the scales of time and velocity are different.
 
\begin{figure}
\plotone{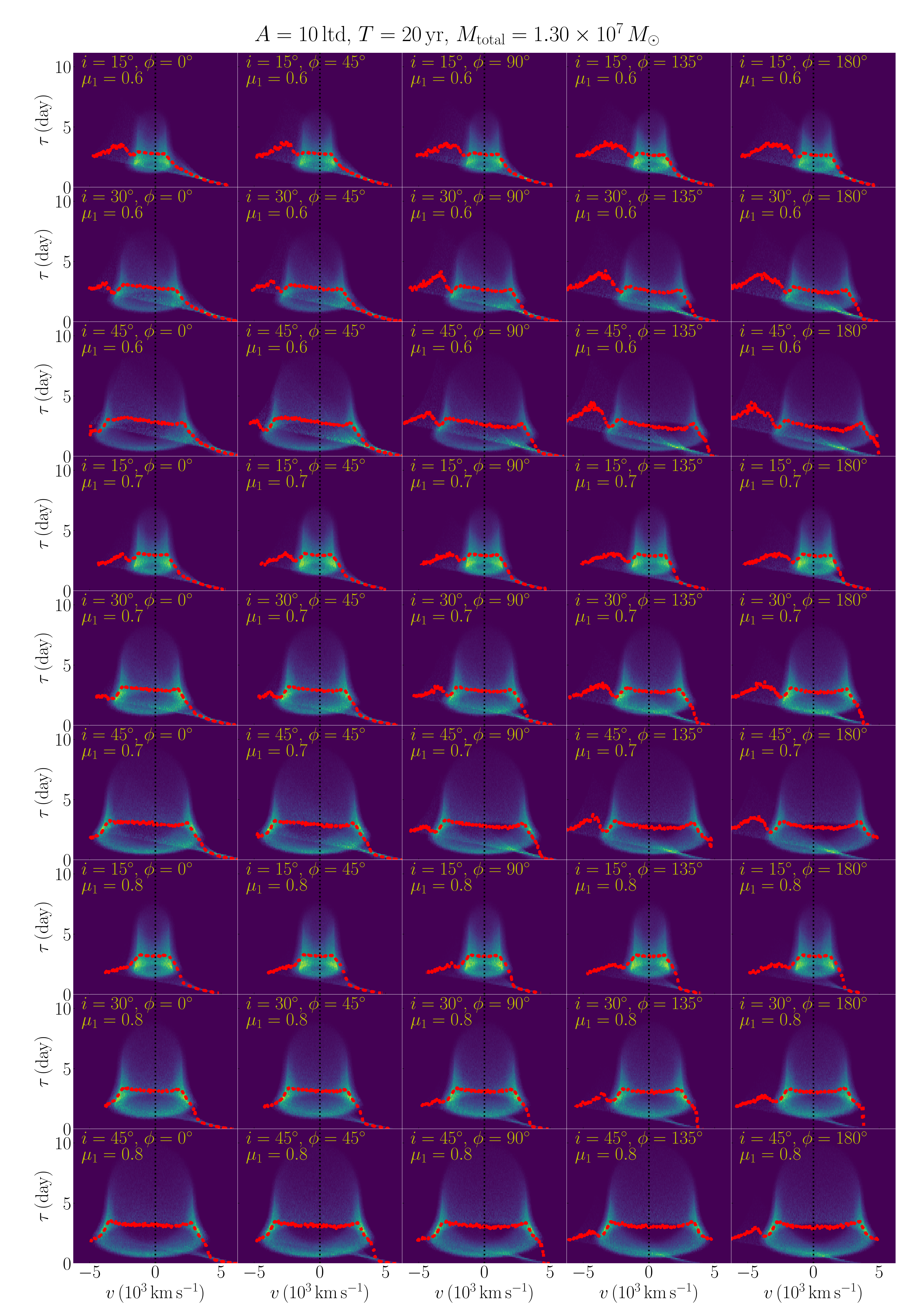}
\caption{\label{fig:trans_inflow} Atlas of 2D TFs of one disk-like BLR and one inflowing BLR with $A = 10$ ltd and $T = 20$ yr.}
\end{figure}

Figure~\ref{fig:trans_inflow} shows the atlas of 2D TFs for a disk-like BLR plus an inflowing BLR.  We only present the case with $A=10$ ltd and $T=20$ yr, since the separation and period only affect the scale of the TFs.  Each 2D TF is a superposition of a bell shape and a fan shape.  The fan shape is inclined so that the delay distribution on the blue side is large and wide, while the delay on the red side is small and narrow.  The impact of inclination, orbital phase, and mass ratio are similar to the case of two disk-like BLRs.  We also note that the velocity-binned time lags show distinctive features when compared to those of a single disk-like BLR.  The blue side of the lags usually shows an extra peak instead of decreasing, while the red side drops to zero.

\begin{figure}
\plotone{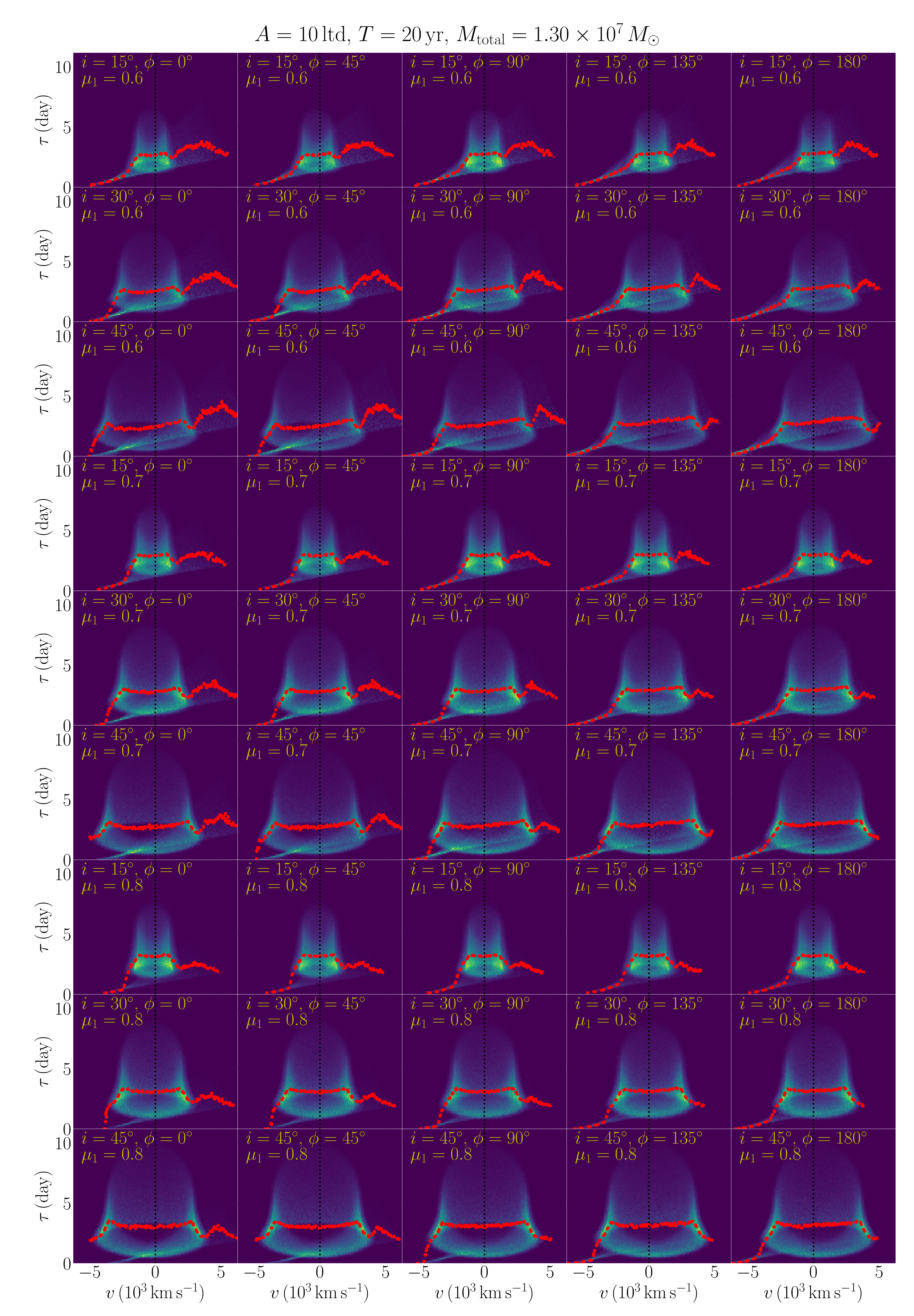}
\caption{\label{fig:trans_outflow} Atlas of 2D TFs of one disk-like BLR and one outflowing BLR with $A = 10$ ltd and $T = 20$ yr.}
\end{figure}

Lastly, the atlas of 2D TFs for a disk-like BLR in combination with an outflowing BLR is shown in Figure~\ref{fig:trans_outflow}.  They are similar to those in Figure \ref{fig:trans_inflow}, except that the fan shape is reflected about the zero velocity line.

\section{Discussion}

We have presented a formulism for calculating the 2D TFs of binary BLRs and 
calculated the TFs using a simple but generic model, with a wide range of 
model parameters. The results are shown as a series of atlases of 2D TFs. 
Given observed TFs from RM campaigns, we can directly compare them with the 
atlases presented here to select candidate CB-SMBHs and roughly infer the 
geometry and kinematics of the constituent BLRs.  Then, detailed analysis, 
such as MCMC, can be applied to obtain the value and uncertainties of model 
parameters.

TFs must be reconstructed from the light curves of continuum and 
velocity-resolved emission lines.  The signal-to-noise ratio (S/N), cadence 
and duration of light curves, and the resolution of the spectra must be 
adequate to allow faithful reconstruction of the TF from RM data and 
comparison with details of the TF predicted from CB-SMBHs.  To address this 
problem, we simulate light curves using a typical CB-SMBH model under 
different observing conditions.  Then, we reconstruct the underlying 2D TFs by 
MEM \citep{Horne1994, Xiao2018a} from the simulated data, and we investigate 
whether the kinematic or geometric features of binary BLRs are preserved for 
candidate selection or identification.

\begin{figure}
\plotone{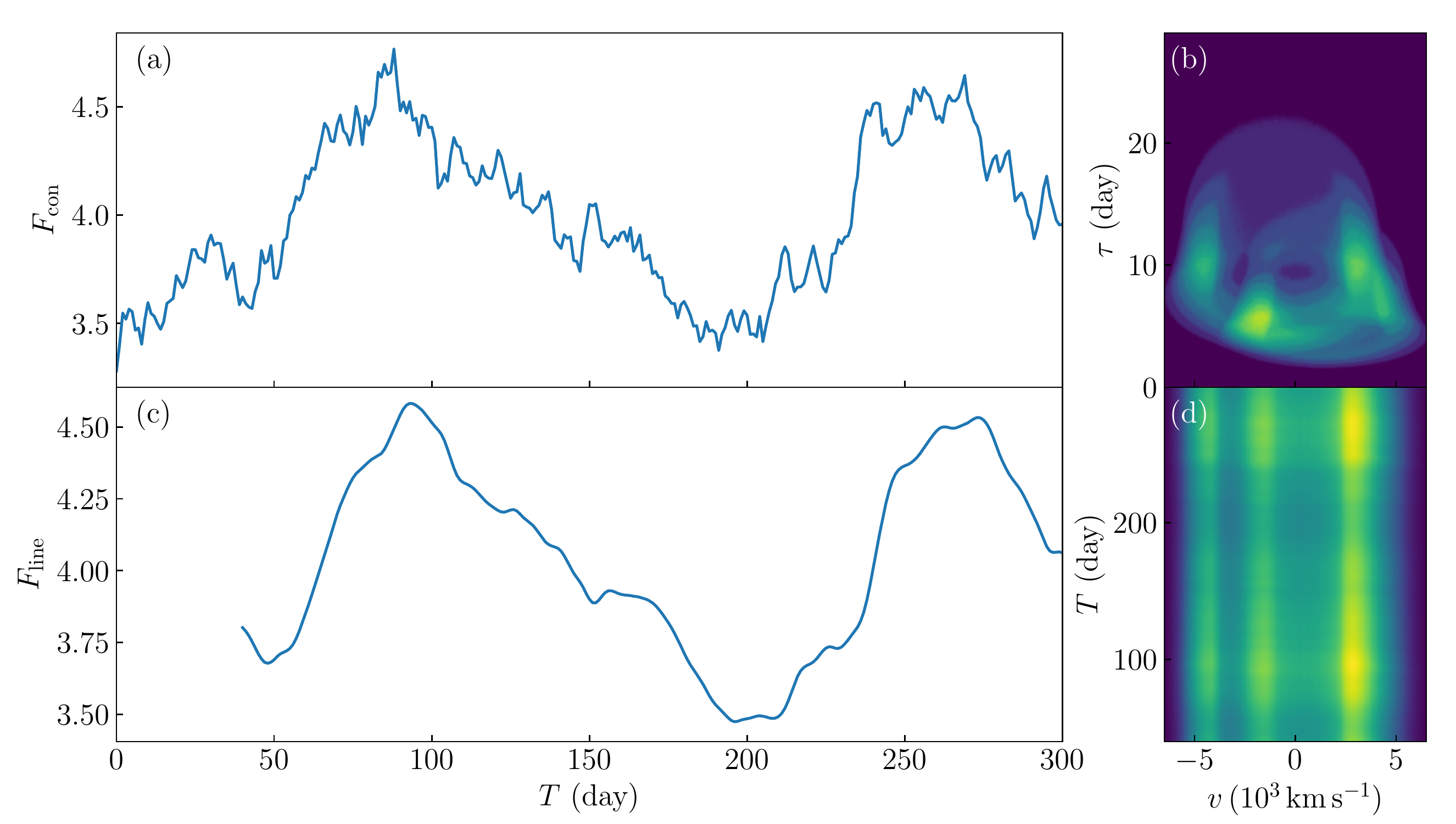}
\caption{\label{fig:simu} Simulated light curves of continuum and emission lines. Panel (a) is the continuum light curve generated by the damped random walk model. Panel (b) is the 2D TF of a typical CB-SMBH. Convolving the continuum light curve with the 2D TF, we obtain the 1D and 2D light curve of the emission line in panels (c) and (d), respectively.}
\end{figure}

We first use the DRW model to generate a continuum light curve with a time span
of 300 days and cadence of 1 day, and then convolve the resulting continuum 
with the model 2D TF of a typical CB-SMBH to get the velocity-resolved 
emission-line light curves.  We assume that the separation between the two BHs 
is $35$ ltd and the period is $38.5$ yr, giving a total BH mass of 
$1.5\times10^8 \, M_{\odot}$.  The BLRs are both flattened Keplerian disks 
with a half-opening angle of $10^{\circ}$.  The inner and outer radii of the 
BLRs are $(7,4)$ ltd and $(15,10)$ ltd, respectively, and the power-law index 
of the reprocessing coefficient distribution for both BLRs is $\gamma = 2$.
The CB-SMBHs are viewed at an inclination of $i = 30^{\circ}$ and an orbital 
phase $\phi = 20^{\circ}$.  The mass fraction of the primary BH is $0.67$.
The simulated RM data with no observational errors and instrumental broadening 
are shown in Figure \ref{fig:simu}.

In the simulations, the uncertainties of the continuum are fixed to $1\%$.
We modify the emission-line profiles to evaluate the influence of two
observational factors:

\begin{description}
\item[Resolution] The spectral resolution is dominated mainly by the 
line-spread function (instrumental broadening). To obtain observed profiles of
different spectra resolutions, we convolve the simulated line profiles with a 
Gaussian broadening function $B(\lambda)$: $L_{\rm \ell,b}(\lambda,t) = 
L_{\ell}(\lambda,t)\otimes B(\lambda)$.  The spectra are sampled with 
intervals equal to half of the FWHM of $B(\lambda)$.

\item[S/N] At every data point $L_{\ell}(v,t)$ of the line profile, a random 
number $\epsilon$ is drawn from a Gaussian distribution with standard 
deviation $\sigma = (S/N)^{-1}$.  The observed noisy data point is then 
$L_{\ell,n} = L_{\ell}(v,t)(1+\epsilon)$.
\end{description}

We enhance the resolution from 1000 to 8000 (by factors of 2) and increase the 
S/N from 25 to 100 (by factors of 2) to generate different simulated RM data.
The MEM-reconstructed 2D TFs from simulated RM data are shown in 
Figure \ref{fig:mem}.  The implementation of MEM is briefly introduced in the 
Appendix.  When the spectral resolution is $\geq 4000$ and the S/N of the 
profile is $\geq 50$, the superposition of the two bell shapes can be seen 
from the reconstructed 2D TF.  There are three discrete bright parts in the 2D 
TF, corresponding to the edge of the bell-shaped TF: the left one is the blue 
side of the Keplerian disk of the primary BH; the middle one is the blue side
of the secondary BLR; and the right one is the overlap of the red sides of the 
two BLRs.  The overlap of the red side indicates that the phase angle is 
between $0^{\circ}$ and $90^{\circ}$.  Compared with the atlas in 
Figure \ref{fig:trans_3_4}, we can conclude that the inclination must be 
larger that $15^{\circ}$, unless all bright parts of the TF are mixed with 
each other.  From the relative size and strength of the two bell shapes, we 
can also infer that the mass fraction of the primary BH must be smaller than
$0.8$.

Clearly, given enough S/N and spectral resolution, it is possible to select 
CB-SMBH candidates by comparing reconstructed 2D TFs from RM data with our 
atlases.  Furthermore, detailed comparison can also indicate the basic 
geometry and kinematics of the binary BLRs and give loose constraints on some 
model parameters.

\section{Conclusions}
We address the problem of whether reverberation mapping can be used to 
identify binary BLRs in CB-SMBHs. 
Given separated BLRs of binary BHs ionized by their own accretion disks,
we demonstrate that the total TF is 
the linear superposition of the individual TFs of two BLRs, so long as the 
continuum light curves can be describe by a damped random walk, and the time 
scale of variation is much shorter than the time span of the light curve but 
longer than the cadence.  When linear superposition holds, reverberation 
mapping of CB-SMBHs can be parameterized by a simple model in which the BLR is 
characterized by a Keplerian, inflowing, or outflowing disk.  We provide 
atlases of 2D TFs for a wide range of geometries and kinematics.
If the spectral resolution is larger than $4000$ and the error is less 
than $2\%$, 2D TFs reconstructed from RM data using MEM can be compared with 
our atlases to select CB-SMBH candidates, constrain the geometry and 
kinematics of the BLRs, and, through detailed MCMC analysis, infer the 
probability distribution of model parameters.

\acknowledgments
We acknowledge the support by National Key R\&D Program of China (grants 2016YFA0400701 and 2016YFA0400702), 
by NSFC through grants {NSFC-11873048, -11833008, -11573026, -11473002, -11721303, -11773029, -11833008, -11690024}, 
and by Grant No. QYZDJ-SSW-SLH007  from the Key Research Program of Frontier Sciences, CAS, 
by the Strategic Priority Research Program of the Chinese Academy of Sciences grant No.XDB23010400. 

\newcommand{\noop}[1]{}

\appendix
\section{Maximum Entropy Method}
MEM is proven to be effective in recovering the 2D TFs from the reverberation signal (e.g., \citealt{Bentz2010, Grier2013, Xiao2018b, Mangham2019}).
We introduce a discrete linearized echo model 
$L_{\ell}(v_{i},t_{k}) = \bar{L}_{\ell}(v_i) + \sum_{j} \Psi(v_{i},\tau_{j})\left[L_{\rm c}(t_{k} -\tau_{j})-\bar{L}_{\rm c}\right]\Delta \tau$ 
to fit the light curve of continuum and velocity resolved emission lines, and to recover the 2D TF $\Psi(v_i,\tau_j)$ simultaneously. 
Here $\bar{L}_{\ell}(v_i)$ is the background spectrum to be fitted,
and $\bar{L}_{\rm c}$ is the referenced continuum level that is fixed to the median of the continuum data. 
The MEM fitting is accomplished by varying the model parameters $\bm{p} = \{\bar{L}_{\ell}(v_i), \Psi(v_{i},\tau_{j}), L_{\rm c}(t_j)\}$ to minimize the quantity $Q={\chi}^{2}-\alpha S$.
Here, ${\chi}^{2}=\sum_{m}\left[D_{m}-\mathscr{M}_{m}(\bm{p})\right]^{2} /\sigma_{m}^{2}$ controls the differences between the data $D_{m}$ and the model prediction $\mathscr{M}_{m}$,
entropy $S=\sum_{n}\left[p_{n}-q_{n}-p_{n}\ln(p_{n}/q_{n})\right]$ is introduced in the MEM fitting to ensure the ``smoothness'' of the model parameters $p_{n}$,
and $q_{n}$ is designed as the ``default value'' of $p_{n}$ and set to weighted averages of ``nearby'' parameters.  For example,
$q(x)=\sqrt{p(x-\Delta x)p(x+\Delta x)}$ for the one-dimensional (1D) models.

MEM has four user-controlled parameters: $\alpha, \calA, \calW$ and $\calWW$, which control the trade-off between
${\chi}^{2}$ and $S$, the aspect ratio of $\Psi(v,\tau)$, and the ``stiffness'' of $\Psi(v,\tau)$ and $L_{\rm c}(t)$,
respectively. Details of the parameter selections can be found in \cite{Xiao2018a}. In our simulation, we fix the values
of $\calA, \calW$ and $\calWW$ to 1 to guarantee the same level of features in $\Psi(v,\tau)$ and $L_{\rm c}(t)$, and vary
the value of $\alpha$ to get similar ${\chi}^{2}$ in the fittings of data with different S/N and spectral resolution.

\begin{deluxetable}{ccl}[htb!]
\tablecaption{Parameters of the binary BLR model \label{tab:par}}
\footnotesize
\tablecolumns{3}
\tablewidth{0pt}
\tablehead{
\colhead{} &
\colhead{Parameters} &
\colhead{Description}
}
\startdata
{     } & $A$ & separation between two BHs \\
{     } & $T$ & rotation period of  the binary system \\
{For the binary} & $\mu_1$ & mass fraction of the primary BH \\
{              } & $i$ & inclination angle of the LOS\tablenotemark{\it a}  \\
{     } & $\phi$  & phase angle of the rotation relative to the LOS\tablenotemark{\it b} \\
{     } & $\Gamma_0$ & coefficients of the linear combination in equation (\ref{tot}) \\
\hline
{     } & $R_{\rm in}$ & inner radius of the BLR \\
{     } & $R_{\rm out}$ & outer radius of the BLR \\
{For individuals\tablenotemark{\it c} } & $\Theta$ & opening angle of the BLR \\
{             } & $\gamma$ & power-law index of clouds/gas distribution \\
{     } & $\alpha_0(\beta_0)$ & velocity\tablenotemark{\it d} of inflow (outflow) at outer radius \\
{     } & $\alpha(\beta)$ & power-law index of inflow (outflow) velocity
\enddata
\tablenotetext{}{$^a$\,$i=0^{\circ}$ indicates face-on orientation.}
\tablenotetext{}{$^b$\,$\phi=0^{\circ}$ indicates that the connection between two BHs is perpendicular to the LOS.}
\tablenotetext{}{$^c$\,There are two sets of these parameters; subscripts are eliminated here.}
\tablenotetext{}{$^d$\,In units of local Keplerian velocity.}
\end{deluxetable}

\begin{figure}
\plotone{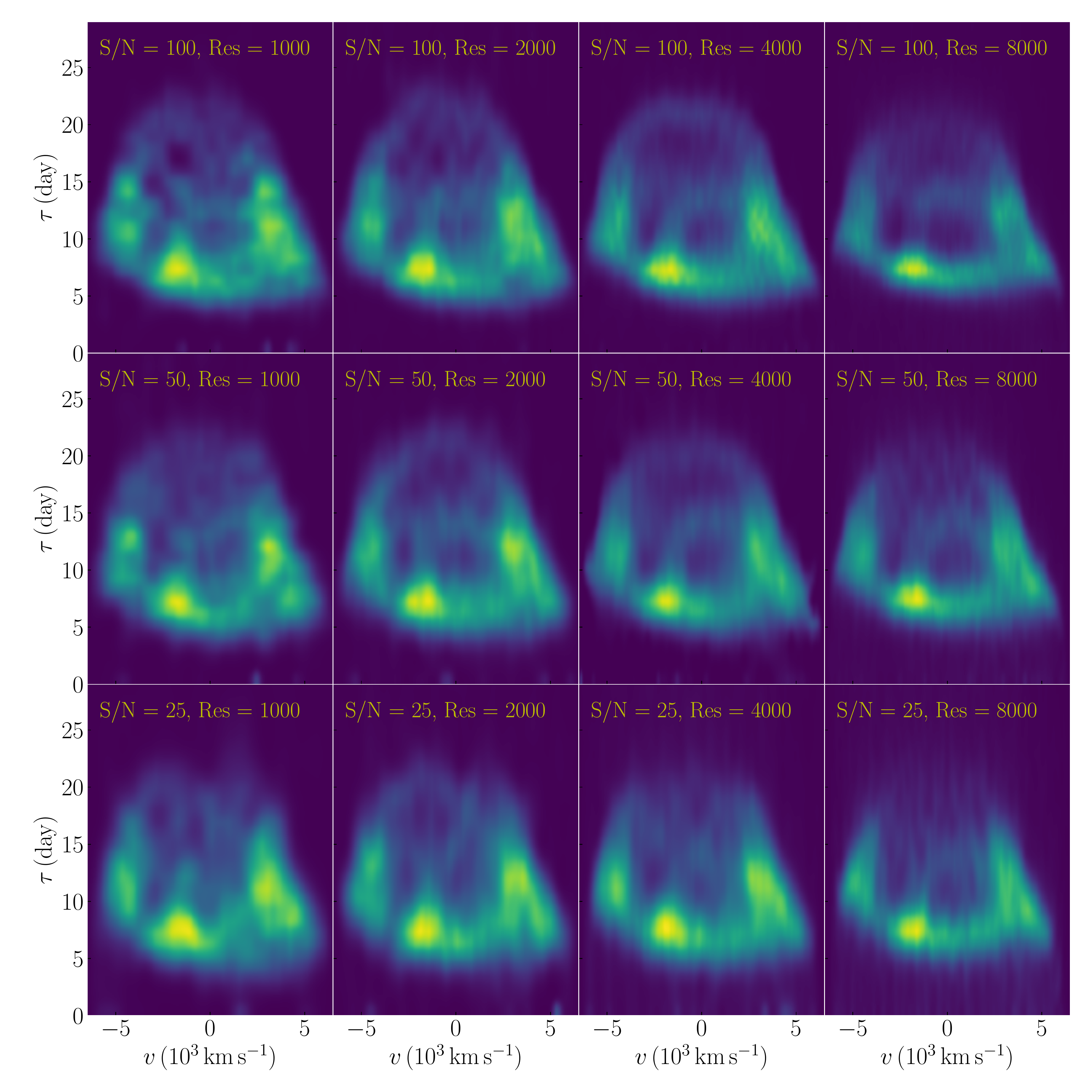} 
\caption{\label{fig:mem} Reconstructed 2D TFs by MEM from simulated data with different S/N and spectral resolution. When the spectral resolution $\ge 4000$ and S/N $\ge 50$, typical features of the CB-SMBH will be present in the reconstructed 2D TF.
}
\end{figure}

\end{document}